\documentclass[american,aps,prl,reprint,longbibliography,superscriptaddress]{revtex4-1}

\usepackage{lineno}
\usepackage{subfigure}
\usepackage{sublabel}
\usepackage{dcolumn}
\usepackage{amsmath,amssymb}
\usepackage{enumitem}

\usepackage{bm}
\usepackage{bbm}
\usepackage{color}
\usepackage{overpic}
\usepackage{latexsym}
\usepackage{epstopdf}
\usepackage{color}
\usepackage[english]{babel}
\usepackage[utf8]{inputenc}
\usepackage{times}
\usepackage{latexsym}
\usepackage{stmaryrd}
\usepackage{psfrag,graphicx}
\usepackage{epsf}
\usepackage{subfigure}
\usepackage{amsmath}
\usepackage{amssymb}
\usepackage{amsfonts}
\usepackage{natbib}
\usepackage{epstopdf}
\usepackage{appendix}
\usepackage{enumitem}
\usepackage{braket} 				

\newcommand{\ketbra}[2]{\vert#1\rangle\!\langle#2\vert}

\def\ket#1{\left|#1\right\rangle}

\definecolor{mygrey}{gray}{0.35}
\definecolor{myblue}{rgb}{0.2,0.2,0.8}
\definecolor{myzard}{cmyk}{0,0,0.05,0}
\definecolor{mywhite}{rgb}{1,1,1}
\definecolor{myred}{rgb}{0.9,0.1,0.}
\usepackage[colorlinks=true,citecolor=myblue,linkcolor=myblue,urlcolor=myblue]{hyperref}

\begin{document}
\title{Experimental Quantum Target Detection Approaching the Fundamental Helstrom Limit}

\author{Feixiang Xu}
\thanks{These two authors contributed equally}
\affiliation{National Laboratory of Solid State Microstructures,
College of Engineering and Applied Sciences and School of Physics, Nanjing University, Nanjing 210093, China}
\affiliation{Collaborative Innovation Center of Advanced Microstructures, Nanjing University, Nanjing 210093, China}

\author{Xiao-Ming Zhang}
\thanks{These two authors contributed equally}
\affiliation{Department of Physics, City University of Hong Kong, Tat Chee Avenue, Kowloon, Hong Kong SAR, China}

\author{Liang Xu}
\affiliation{National Laboratory of Solid State Microstructures,
College of Engineering and Applied Sciences and School of Physics, Nanjing University, Nanjing 210093, China}
\affiliation{Collaborative Innovation Center of Advanced Microstructures, Nanjing University, Nanjing 210093, China}

\author{Tao Jiang}
\affiliation{National Laboratory of Solid State Microstructures,
College of Engineering and Applied Sciences and School of Physics, Nanjing University, Nanjing 210093, China}
\affiliation{Collaborative Innovation Center of Advanced Microstructures, Nanjing University, Nanjing 210093, China}

\author{Man-Hong Yung}
\email{yung@sustech.edu.cn}
\affiliation{Department of Physics, Southern University of Science and Technology, Shenzhen 518055, China}
\affiliation{Shenzhen Institute for Quantum Science and Engineering, Southern University of Science and Technology, Shenzhen 518055, China}
\affiliation{Guangdong Provincial Key Laboratory of Quantum Science and Engineering, Southern University of Science and Technology, Shenzhen 518055, China}
\affiliation{Shenzhen Key Laboratory of Quantum Science and Engineering, Southern University of Science and Technology, Shenzhen,518055, China}

\author{Lijian Zhang}
\email{lijian.zhang@nju.edu.cn}
\affiliation{National Laboratory of Solid State Microstructures,
College of Engineering and Applied Sciences and School of Physics, Nanjing University, Nanjing 210093, China}
\affiliation{Collaborative Innovation Center of Advanced Microstructures, Nanjing University, Nanjing 210093, China}

\begin{abstract}
Quantum target detection is an emerging application that utilizes entanglement to enhance the sensing of the presence of an object. Although several experimental demonstrations for certain situations have been reported recently, the single-shot detection limit imposed by the Helstrom limit has not been reached because of the unknown optimum measurements. Here we report an experimental demonstration of quantum target detection, also known as quantum illumination, in the single-photon limit. In our experiment, one photon of the maximally entangled photon pair is employed as the probe signal and the corresponding optimum measurement is implemented at the receiver. We explore the detection problem in different regions of the parameter space and verify that the quantum advantage exists even in a forbidden region of the conventional illumination, where all classical schemes become useless. Our results indicate that quantum illumination breaks the classical limit for up to $40\%$, while approaching the quantum limit imposed by the Helstrom limit. These results not only demonstrate the advantage of quantum illumination, but also manifest its valuable potential of target detection.
\end{abstract}
\maketitle

{\em Introduction--}
Using non-classical resources for practical applications is an intriguing topic in quantum information processing. An entanglement-based protocol for quantum target detection under background noise, named quantum illumination, has been developed in the past decade~\cite{lloyd2008enhanced,PhysRevLett.101.253601,5205594,Shapiro_2009,PhysRevA.79.062320,PhysRevA.80.052310,Ragy:14,PhysRevLett.118.040801,PhysRevLett.118.070803,PhysRevLett.119.120501,Weedbrook_2016,PhysRevA.95.022333,PhysRevLett.110.153603,PhysRevA.96.020302,yang2017optomechanical,PhysRevLett.114.110506,PhysRevLett.114.080503,lee2020quantum,karsa2020noisy,barzanjeh2020microwave,Zhang_2020} that can potentially be applied to many areas, including quantum secure communication \cite{PhysRevLett.111.010501,shapiro2014secure,PhysRevA.80.022320}, quantum sensing \cite{las2017quantum}, quantum imaging \cite{PhysRevA.99.023828,Gregoryeaay2652,Deenne2020quantum}, quantum reading \cite{PhysRevLett.106.090504,PhysRevA.87.062310,PhysRevA.86.012315} and quantum radar \cite{chang2019quantum,luong2019receiver,jonsson2020comparison}.

The advantages of quantum illumination have been theoretically investigated for different probe states~\cite{PhysRevLett.101.253601,PhysRevA.79.062320,PhysRevA.89.062309,PhysRevA.95.042317,PhysRevA.98.012319,Zhang_2020} and the joint measurement methods \cite{5205594,PhysRevA.80.052310,PhysRevLett.118.040801,Jo2020quantum}.
Some of these quantum illumination schemes have been experimentally demonstrated~\cite{PhysRevLett.110.153603,PhysRevLett.111.010501,PhysRevLett.114.110506,PhysRevA.101.053808,PhysRevX.9.031033,barzanjeh2020microwave,PhysRevA.99.053813}. However, there are few studies on the ultimate limit of the quantum illumination. Although several protocols have been proposed to reach the quantum limit \cite{PhysRevA.100.012327,pirandola2019fundamental,ranjith2020fundamental}, to our best knowledge, no experiment has been demonstrated yet, impeding the maximum gain in terms of utilizing quantum resources. Moreover, a comparison between the performance of the quantum illumination and the fundamental limit of the conventional illumination, which is crucial for the demonstration of the unambiguous quantum advantage,  has rarely been studied experimentally.

Here, we experimentally demonstrate a single-photon quantum illumination scheme that can approach the fundamental limit imposed by the Helstrom limit. In our experiment, one photon of the maximally entangled photon pair is used as the probe signal state sent to the target.
The corresponding optimum measurement is implemented at the receiver. We experimentally explore different regions in the parameter spaces, where the error probabilities of different reflectivity ranging from 0 to 1 of the target are investigated, not limited to the small reflectivity like most of the illumination schemes. The experimentally measured error probabilities for quantum illumination are compared to the classical limit in the whole parameter space.

In particular, for both conventional and quantum illuminations, we identify two different subregions called the forbidden regions (where the performance of detection for any probe signal cannot surpass the direct guess) and the illuminable regions (where the detectability can be optimized by varying the signal states and the measurement apparatus).
In illuminable region, compared to the optimum conventional illumination, the quantum illumination breaks the classical limit for up to 40 $\%$, while approaching the fundamental limit with a small discrepancy due to experimental imperfections. Even in the forbidden region of the conventional illumination, the quantum advantage in terms of the detection error is experimentally observed. Additionally, the quantum advantage is also quantified in terms of the mutual information, which is in good agreement with the theoretical prediction.

\begin{figure}[t]
	 \centering
   \includegraphics[width=0.47\textwidth]{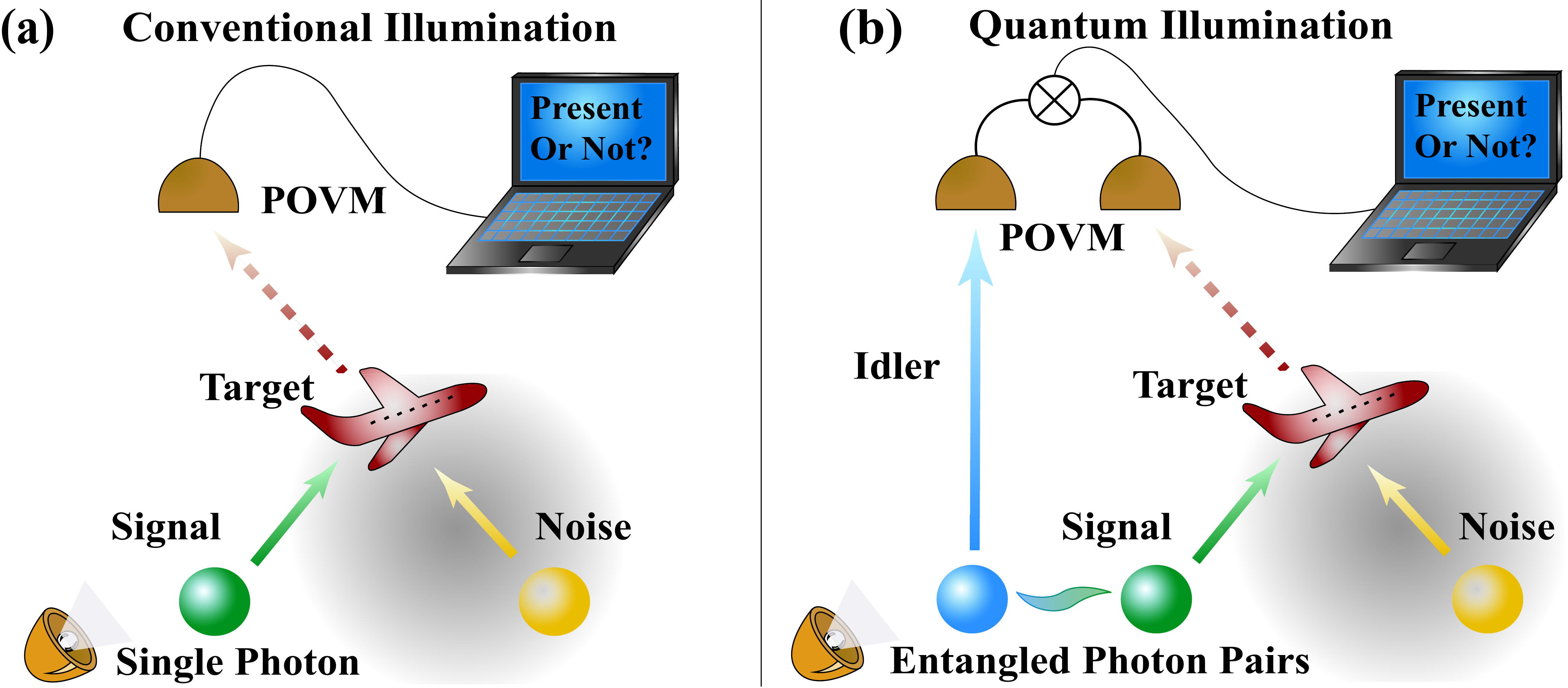}
   \caption[width=1\textwidth]{A schematic of (a) conventional and (b) quantum illumination. In conventional illumination (a), a single photon as a probe signal state is sent to a suspected background noise region. An optimum positive operator-valued measure (POVM) is designed to detect the potential object. In quantum illumination (b), a pair of entangled photon is prepared, one of which (the idler) is directly sent to the receiver. The other one (the signal) is sent to the suspected region for illumination. A joint measurement is implemented at the receiver to detect the potential object.}
	\label{fig:schematic}
\end{figure}

{\em Background}--The schematics \cite{lloyd2008enhanced,yung2020one} of the conventional and quantum illumination in the single-photon limit are shown in Fig.\ref{fig:schematic}. In conventional illumination, a single photon $\hat{\rho}_c$ is used as a probe signal. If the target is absent, $\hat{\rho}_c$ is completely lost and the detectors can only receive the background noise $\hat{\rho}_E$, i.e., $\hat{\rho}_c^{(0)}=\mathcal{E}_0(\hat{\rho})=\hat{\rho}_E$. If the target is present,
the received state becomes $\hat{\rho}_c^{(1)}=\mathcal{E}_1(\hat{\rho})=\eta\hat{\rho}_c+(1-\eta)\hat{\rho}_E$, with the mixing ratio $\eta\in [0,1]$. In quantum illumination, the strategy can be improved by resorting to the quantum entanglement.
One photon of the entangled photon pair $\hat{\rho}_{SI}$
is used as probe signal state to illuminate the target, and the other one serves as an idler retained for future measurement. When the target is absent, the received bipartite quantum state is the direct tensor product of the background state and the idler state $\hat{\rho}_q^{(0)}=(\mathcal{E}_0\otimes\mathcal{I})(\hat{\rho}_{SI})=\hat{\rho}_E\otimes\hat{\rho}_I$,
where $\hat{\rho}_I=\text{Tr}_S(\hat{\rho}_{SI})$ is the partial trace of $\hat{\rho}_{SI}$ over the signal photon. If the target is present, the received state becomes $\hat{\rho}_q^{(1)}=(\mathcal{E}_1\otimes\mathcal{I})(\hat{\rho}_{SI})=\eta\hat{\rho}_{SI}+(1-\eta)\hat{\rho}_E\otimes\hat{\rho}_I$.

For fixed probe states, discriminating the channel $\mathcal{E}_1$ from $\mathcal{E}_0$ is equivalent to discriminating a pair of quantum states $\hat{\rho}_{c/q}^{(0)}$ and $\hat{\rho}_{c/q}^{(1)}$. We denote $p_0$ and $p_1$ as the prior probabilities of presence or absence of the target, where $p_0+p_1=1$. By optimizing the quantum measurement, the error probability of discriminating $\hat{\rho}_{c/q}^{(0)}$ and $\hat{\rho}_{c/q}^{(1)}$ for single-shot measurement
is lower bounded by the Helstrom limit
\cite{helstrom1969quantum}
\begin{equation}
  P^{\text{HL}}_\text{err}\equiv\frac{1}{2}(1-||p_0\hat{\rho}_{c/q}^{(0)}-p_1\hat{\rho}_{c/q}^{(1)}||),
  \label{eq:Helstrom}
\end{equation}
where $||A||\equiv\text{tr}\sqrt{A^\dagger A}$ denotes the trace norm of $A$. Our ultimate task is to minimize the error probability $P^{\text{HL}}_\text{err}$ over all possible quantum states, under given prior probabilities $p_0, p_1$ and the mixing ratio $\eta$.

{\em Protocol--} The states that can achieve the ultimate illumination strategy depend only on the spectral information of the background noise~\cite{yung2020one}, which can generally be a mixed states described by $\hat\rho_E=\sum_{i=0}^{d-1}\lambda_i\ketbra{i}{i}$ with $\lambda_i$ ordered in an ascending way. For conventional illumination, the optimum state should be $\hat\rho_c=\ketbra{0}{0}$ corresponding to the eigenstate of $\hat{\rho}_E$ with minimum eigenvalue. For quantum illumination, the optimum state should be
 $\hat\rho_{SI}=\ketbra{\psi_{AB}}{\psi_{AB}}$, where $\ket{\psi_{AB}}=\sum_{i=0}^{d-1}\mu_i\ket{i}\ket{i}$
with $\mu_i=\sqrt{\frac{1}{(\sum_{j=0}^{d-1}\lambda_j^{-1})\lambda_i}}$. According to Helstrom's theorem, the nontrivial optimum two-outcome measurement operators $\{\hat{\Pi}_{c/q}^{(0)},\hat{\Pi}_{c/q}^{(1)} \}$
exist only for the parameter regions that allow the existence of both the positive and negative eigenvalues of operators $\hat{\Omega}_{c/q}=p_0\hat{\rho}_{c/q}^{(0)}-p_1\hat{\rho}_{c/q}^{(1)}$, named the “illuminable region”. Otherwise,
we say the parameters are in the “forbidden region”, in which the measurement operators are trivial so that one can achieve the minimum error probability by direct guess. In the illuminable region, the optimum measurement operators are proved to be
\begin{equation}
    \hat{\Pi}_{c}^{(1)}=\ketbra{0}{0},\;    \hat{\Pi}_{c}^{(0)}=\mathbb{I}_d-\hat{\Pi}_{c}^{(1)}
  \label{eq:conmeasure}
\end{equation}
for conventional illumination, and
\begin{equation}
     \hat{\Pi}_q^{(1)}=\ketbra{\psi_{AB}}{\psi_{AB}},\;  \hat{\Pi}_q^{(0)}=\mathbb{I}_d\otimes\mathbb{I}_d-\hat{\Pi}_q^{(1)}
  \label{eq:quanmeasure}
\end{equation}
for quantum illumination \cite{yung2020one}. Here, we use $\mathbb{I}_d$ to denote the $d$-dimensional identity operator. The absence or presence of the target can be decided according to the measurement outcome with minimum error probability: If the received states are projected on $\hat{\Pi}^{(0)}_{c/q}$, one can declare the absence of the target. Otherwise, one can guess the target is present.

The above protocol is general and can describe any systems with photon pairs correlated in different degrees of freedoms, e.g., the spatial correlations \cite{Gregoryeaay2652}, spectro-temporal correlations \cite{PhysRevA.101.053808}. In this work, we make use of the polarization-entangled photon pairs to achieve the advantages of quantum illuminations. Under polarization degrees of freedom, the background noise is natural to be $\hat{\rho}_E=\frac{1}{2}\ketbra{H}{H}+\frac{1}{2}\ketbra{V}{V}$. Therefore, the optimum states should be the maximally polarization-entangled state $\ket{\Phi^+}=1/\sqrt{2}(\ket{HH}+\ket{VV})$.
The corresponding optimum measurement operators should be $\hat{\Pi}_q^{(1)}=\ketbra{\Phi^+}{\Phi^+}$ and $\hat{\Pi}_q^{(0)}=\mathbb{I}\otimes\mathbb{I}-\hat{\Pi}_q^{(1)}$, which is the Bell state measurement (BSM) \cite{PhysRevA.57.2208}.

\begin{figure*}[t]
	 \centering
   \includegraphics[width=1\textwidth]{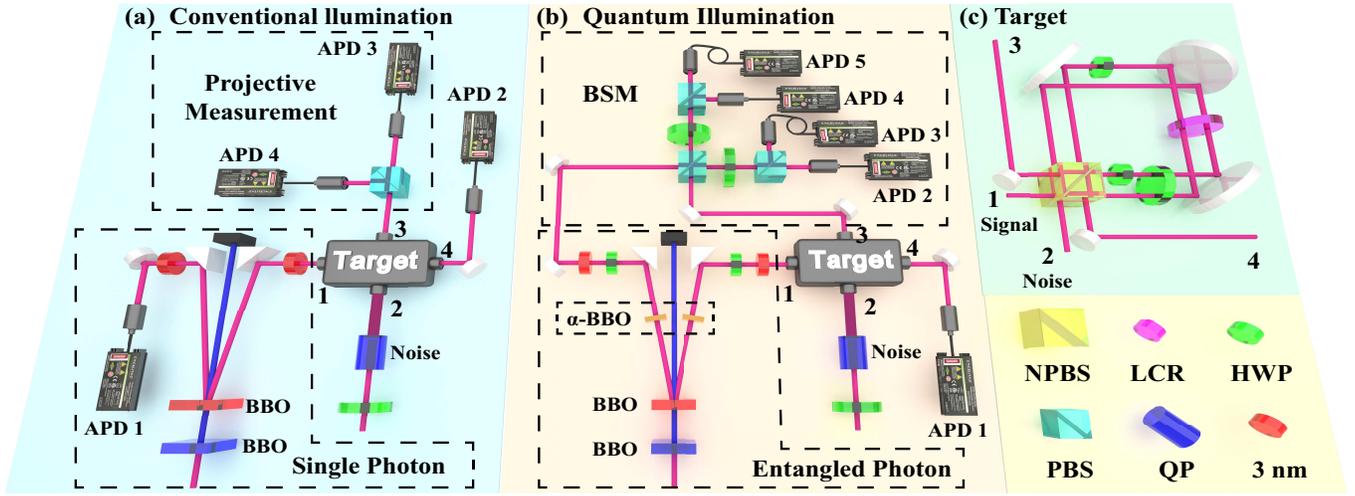}
   \caption[width=1\textwidth]{Experimental setup. (a) Conventional illumination: The signal is a heralded single photon, which is sent into port 1 of the target for illumination. After combination with the background noise from port 2, the output of port 3 is sent to the projective measurement for deciding the absence or presence of the target. (b) Quantum illumination: The heralded single photon is replaced with a photon entangled with an idler, which is sent to the Bell state measurement (BSM) directly. The signal is combined with the background noise and the output of port 3 is sent to the BSM for the joint measurement with the idler. (c) Target: a Sagnac interferometer-based polarization-independent beam splitter with tunable reflectivity. Key elements included half wave plate (HWP), PBS (polarizing beam splitter), 3 nm (3 nm bandpass filter), QP (quartz plate), NPBS (nonpolarizing beam splitter), APD (avalanche photodiode)}
	\label{fig:experiment}
\end{figure*}

{\em Experiment--} To verify the validity of the results, we implement both conventional [Fig.\ref{fig:experiment}(a)] and quantum illumination [Fig.\ref{fig:experiment}(b)] experiments at all  parameter regions, where the number of photons used to detect the target is the same. To change the reflectivity $R$ flexibly, we use a a tunable polarization-independent beam splitter implemented by displaced Sagnac interferometer \cite{ried2015quantum,nagata2007beating,kwiat2000grover} as the target [Fig.\ref{fig:experiment}(c)]. In this way, the reflectivity $R$ can be adjusted conveniently by inserting the appropriate phase difference between the two displaced paths in the interferometer (see~\cite{Supplementary_material} for details). When the reflectivity $R=0$, the target is absent. The prior probability $p_0$ is set if the trials ratio of $R=0$ to the whole experiment trials is $p_0$. The error rate is then averaged over all experimental trials.

For conventional illumination [Fig.\ref{fig:experiment}(a)], we make use of the signal photon to illuminate the tunable target, while the idler photon is used as heralding. The projective measurement is then implemented at the receiver to perform the optimum measurement. For quantum illumination, the heralded single-photon source is replaced with a polarization-entangled photon state having fidelity 96.4\% with the Bell state $\ket{\Phi^+}=\frac{1}{\sqrt{2}}(\ket{HH}+\ket{VV})$. One photon of the entangled photon pair as a signal is sent to the target for illumination, while the other one, as an ancilla, is directly sent to the joint measurement apparatus, the BSM. If there is twofold coincidence between the two single-photon detectors APD 2 and APD 5 or APD 3 and APD 4, the ancilla together with the photon from the target are projected on the maximally entangled state $\ket{\Phi^+}$ and we guess the target is present. Otherwise, the outcome of the apparatus corresponds to the measurement operator $\mathbb{I}-\ketbra{\Phi^+}{\Phi^+}$, and we guess the target is not present.
More details of the experimental setup can be found in the Supplementary Material \cite{Supplementary_material}

\begin{figure}[t]
	 \centering
   \includegraphics[width=0.45\textwidth]{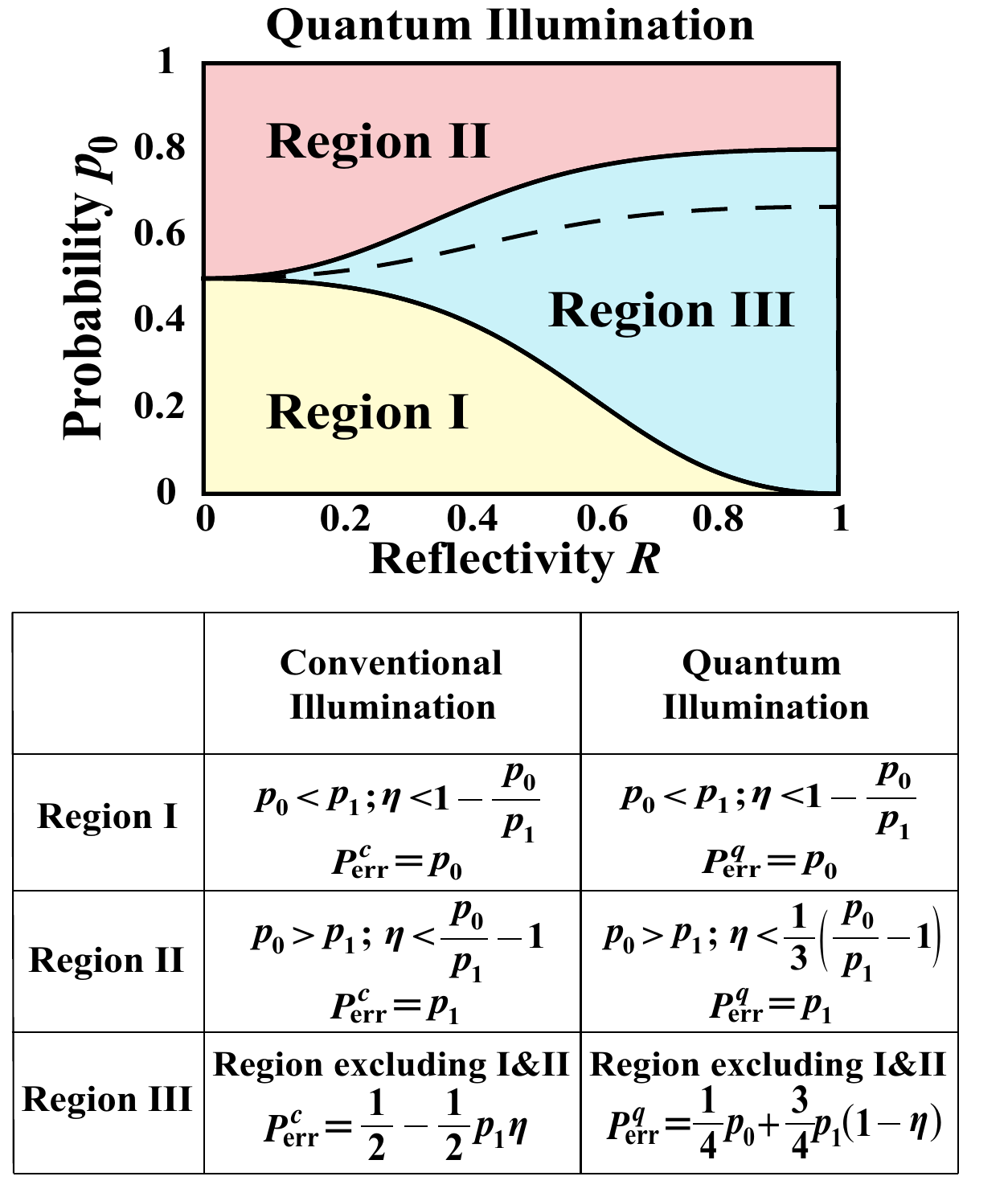}
   \caption[width=1\textwidth]{Phase diagram for quantum illumination. The black solid lines denote the boundaries between different regions. The black dashed line represents the boundary between region II and region III of conventional illumination in the quantum illumination phase diagram for comparison. The table below the phase diagram shows the range of the three regions and their minimum error probabilities, where $\eta=R^2/(R^2+(1-R)^2)$}
	\label{fig:simulation}
\end{figure}

{\em Results}--
The prior probability of absence of the target $p_0$ and its reflectivity $R$ expand a parameter space, forming a ``phase diagram'' with three regions according to the theoretical minimum error probability. Both region I and region II are the forbidden regions, where the minimum error probability is achieved by the trivial direct guess. The illuminable region for both illumination schemes is region III, in which the corresponding detection limits are achieved according to the measurement outcomes. In the ``phase diagram'' for quantum illumination shown in Fig.~\ref{fig:simulation}, the black dashed line in region III denotes the bound of region II and III for conventional illumination~\cite{yung2020one}. Clearly, the illuminable region for quantum illumination is larger, indicating that quantum illumination for the parameters in the larger region can break the classical limit obtained by the trivial direct guess. The table below the phase diagram shows the ranges of the parameters for three regions and their minimum error probabilities. Although the mixing probability $\eta$ of the phase diagram in Ref.~\cite{yung2020one} can be equivalently calculated with the reflectivity $R$, i.e., $\eta=R^2/(R^2+(1-R)^2)$ \cite{Supplementary_material}, it is more natural to investigate the problem of the target detection with the parameter reflectivity $R$ as the intrinsic property of the target.

\begin{figure}[t]
	 \centering
   \includegraphics[width=0.47\textwidth]{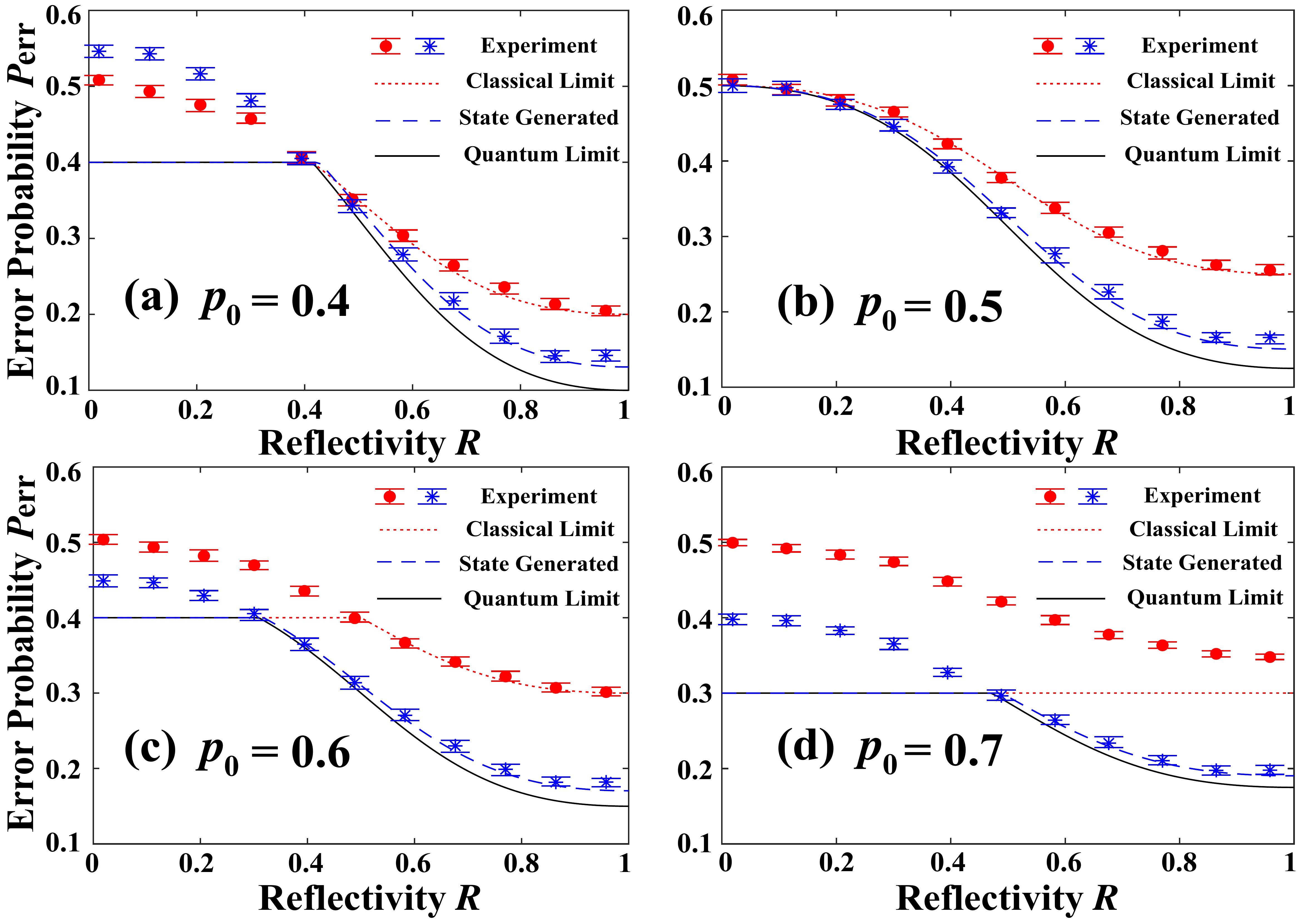}
   \caption[width=1\textwidth]{Experimentally measured error probabilities of conventional (red dots with error bars) and quantum (blue asterisk with error bars) illuminations for different prior probabilities (a) $p_0=0.4$, (b) $p_0=0.5$, (c) $p_0=0.6$ and (d) $p_0=0.7$. The theoretical classical and quantum limit are shown as red dotted lines and black solid lines, respectively. The theoretical error probabilities for quantum illumination utilizing the state generated in the experiment are depicted as the blue dashed lines.}
	\label{fig:results}
\end{figure}

To verify these results, in our experiment, we choose four different prior probabilities $p_0=0.4,~0.5,~0.6$ and 0.7 to experimentally measure the corresponding error probabilities, which are illustrated in Fig.\ref{fig:results} by dots (conventional) and asterisk (quantum) and the error bars originate from the statistical fluctuations. For comparison, we also plot the theoretical minimum error probabilities for both conventional (red dotted lines) and quantum (black solid lines) illuminations. It can be seen that for some values of the reflectivity, the predicted minimum error probabilities are constants in Fig.\ref{fig:results}(a), (c) and (d). The reason for this phenomena is that apart from the prior probability $p_0=0.5$, the error rates for other prior probabilities go through not only the illuminable region, but also the forbidden region (region I or region II), in which the minimum error probabilities depend only on the prior probability $p_0$.

In the illuminable region where the predicted minimum error probabilities are not constants, the experimentally measured error probabilities for conventional illumination (red dots with error bars) coincide well with the theoretical predictions. For quantum illumination, the detected error probabilities are only $9\%$ away from the fundamental quantum limit (black solid lines) imposed by Helstrom limit on average, and even $0.1\%$ for several instances.
Even so, quantum illumination can still reduce the detection error up to 40 $\%$ compared to conventional illumination.

To better understand 9\% discrepancy between the quantum Helstrom limit and the experimental results, we characterize the probe state generated in the experiment and calculate the expected error probabilities with this state, the results of which are shown as a blue dashed line in Fig.~\ref{fig:results}. To rigorously quantify the contributions of the imperfect quantum states to the discrepancy, we estimate the relative ratio of the difference between the error rates derived from the imperfect states generated in the experiment and the quantum limit to the difference between the experimentally measured error probabilities and the quantum limit. The results indicate that about 80\% of the discrepancy between the experimental results and the Helstrom limit comes from the imperfections in the probe states. The remaining errors may originate from other error sources such as photon loss and imperfect measurement apparatus. If we utilize the ideal Bell state as the probe state and perform the perfect BSM, our quantum illumination scheme can achieve the Helstrom limit and the maximum improvement, which should be 50\% compared to the classical limit.

In the forbidden region, however, no illumination schemes can achieve lower error probabilities than the direct guess. If we insist on determining the presence or not of the target according to the outcomes of the optimum measurement operators of the illuminable region, we can see from the Fig.~\ref{fig:results}(a),(c) and (d) that the error probabilities are higher than the direct guess. The advantages of the measurements implemented for illuminable region thus disappear rapidly. Yet the area of the forbidden area is smaller for quantum illumination. From the results shown in Fig.~\ref{fig:results}(c) and (d), the experimental results demonstrate that for certain values of $R$, quantum illumination can beat direct guess while conventional illumination cannot. In particular, for $p_0=0.7$, conventional illumination is useless for any values of $R$, but quantum illumination can still provide enhancement when $R>0.5$, which confirms the advantage of quantum illumination in shrinking the forbidden region.

\begin{figure}[t]
	 \centering
   \includegraphics[width=0.4\textwidth]{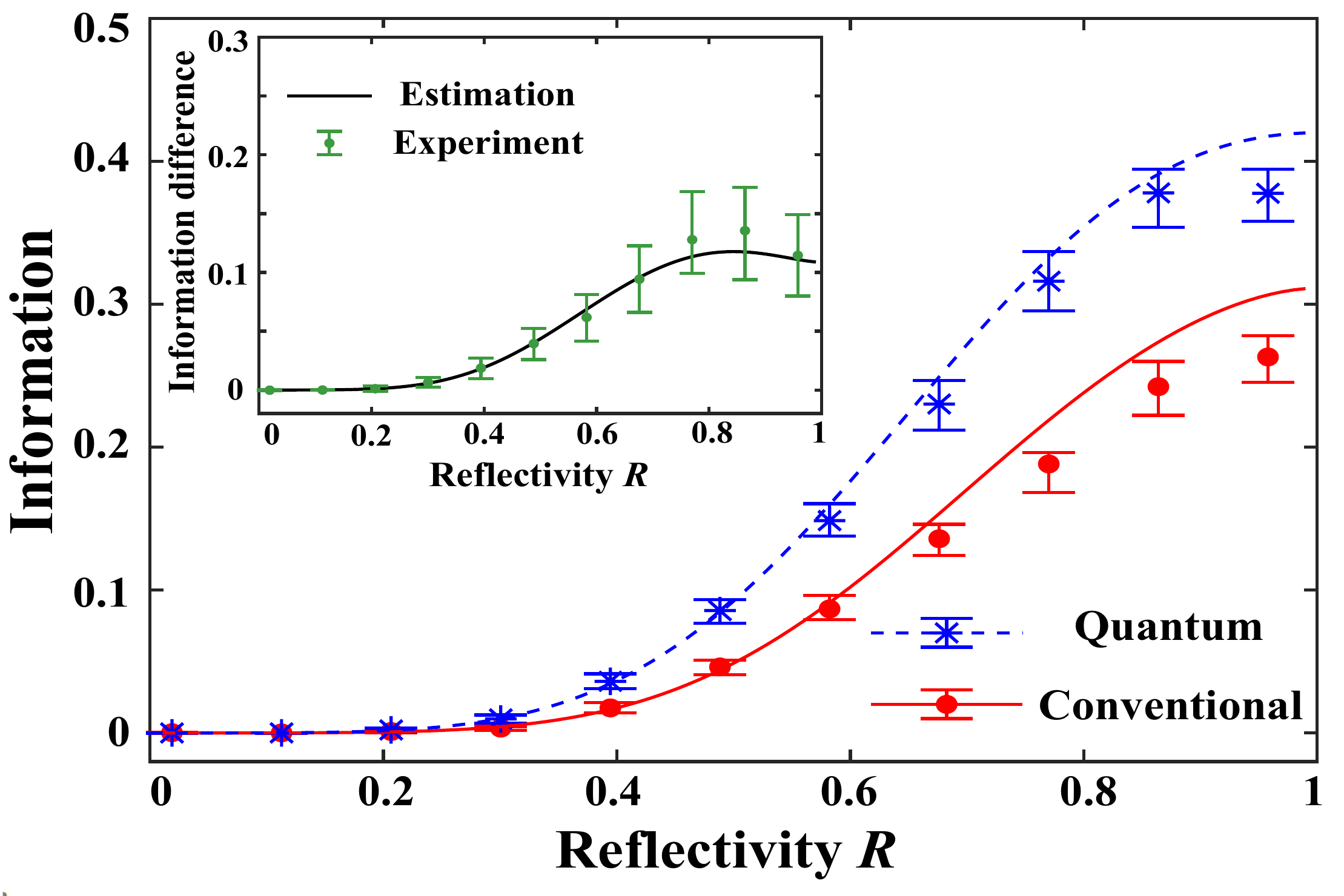}
   \caption[width=1\textwidth]{Mutual information of quantum and conventional illumination and their difference (inset). The red solid line and the blue dashed line indicate the theoretical mutual information of conventional $I_c$ and quantum $I_q$ illumination utilizing the state generated from the experiment with respect to $p_0=0.5$. The red dots and blue asterisk with error bars are their corresponding experimental results. The inset shows the difference of the mutual information.}
	\label{fig:in}
\end{figure}

As suggested in Ref.~\cite{Weedbrook_2016}, the quantum advantage can also be quantified by the difference of the mutual information of quantum $I_q$ and conventional illumination $I_c$ where $I_{c/q}=\max_{\hat{\Pi}} I(p_0,p_{\hat{\Pi}})$ is the mutual information maximized over all the measurement operators with respect to the prior probability $p_0$ and the input states. In Fig.~\ref{fig:in} we show the experimentally estimated mutual information and their difference (inset) for $p_0 = 0.5$.
The experimental results agree well with the theoretical estimation, which further support optimality of our protocol.

{\em Conclusions and Outlook--}
In summary, we demonstrate a quantum illumination scheme that can approach the quantum limit for single-shot detection. We experimentally prepare the maximally entangled state as the optimum probe states to illuminate the potential target. The error probability is obtained experimentally in different parameter regions, which is compared to the classical limit, verifying the unambiguous quantum advantage. Moreover, the experimental results clearly demonstrate the advantage of the quantum illumination reaching the detection limit imposed by the Helstrom limit.

Future extension of quantum illumination along this direction may cover the situation of multishot detection limits, where the forbidden regions are expected to be reduced. However, for these cases, the quantum illumination loses its advantage asymptotically if one simply repeat the optimal detection method in single-photon cases \cite{yung2020one}. The theoretical limit is still under investigation in the literature. On the other hand, our results may also be of interest from the point of view of ``hiding" a certain object from being detected. By restricting the occurrence probability to stay inside the forbidden regions, it becomes ``safe" as no signal, entangled or not, can reveal extra information about the occurrence events relative to blind guessing.

\begin{acknowledgments}
  We thank Xian-Min Jin and Ruo-Jing Ren for helpful advice on the construction of the entangled photon source. This work was supported by the National Key Research and Development Program of China under (Grant Nos. 2017YFA0303703 and 2019YFA0308700), the National Natural Science Foundation of China (Grant No. 61975077, No. 91836303, No. 11690032, No. 11875160 and No. U1801661), Fundamental Research Funds for the Central
  Universities (Grant No. 021314380197, No. 020414380175), Natural Science Foundation of Guangdong Province (Grant No. 2017B030308003), the Key-Area R$\&$D Program of Guangdong province (Grant No. 2018B030326001), the Science, Technology and Innovation Commission of Shenzhen Municipality (Grant Nos. JCYJ20170412152620376, JCYJ20170817105046702 and KYTDPT20181011104202253), the Economy, Trade and Information Commission of Shenzhen Municipality (Grant No. 201901161512), Guangdong Provincial Key Laboratory (Grant No. 2019B121203002).
\end{acknowledgments}
%

\end{document}